\documentclass[11pt]{article}

\usepackage{epsfig}
\usepackage{graphicx}

\begin{document}
\thispagestyle{empty}    

%\begin{titlepage}
\begin{flushright}
UCLA/03/TEP/12\\
UOSTP-03102\phantom{abcd}\\
{\tt hep-th/0304129}\phantom{ab}
\end{flushright}
\vspace{.3cm}

\renewcommand{\thefootnote}{\fnsymbol{footnote}}
\centerline{\Large \bf 
A Dilatonic Deformation of $AdS_5$ and its Field Theory Dual
}

\vskip 1.2cm  
\centerline{ Dongsu Bak,$\!^1$\footnote{email: dsbak@mach.uos.ac.kr}
 Michael  Gutperle$^{2,3}$\footnote{email: gutperle@physics.ucla.edu} and
Shinji Hirano$^4$\footnote{email: hirano@physics.technion.ac.il}
}

\vskip 10mm   
\renewcommand{\thefootnote}{\arabic{footnote}}
\setcounter{footnote}{0}
%\vfill%\vskip 7mm%1cm

\centerline{$^1$ 
Physics Department, University of Seoul, Seoul 130-743, Korea}
\vskip 3mm    
\centerline{ $^2$
Department of Physics and Astronomy, UCLA, Los Angeles,
CA, USA\footnote{Permanent address}}
\vskip 3mm   
\centerline{$^3$ ITP, Department of Physics, Stanford
University, Stanford, CA, USA}
\vskip 3mm   
\centerline{$^4$ Department of Physics, Technion, Israel Institute of
  Technology, Haifa 32000, Israel 
\phantom{abcdefg}} 
\vspace*{0.4cm}
\vskip 20mm

\baselineskip 18pt

%\begin{center}
%{\bf Abstract}
%\end{center}
\begin{quote}
We find a nonsupersymmetric 
dilatonic deformation of $AdS_5$ geometry as
an exact nonsingular solution of the type IIB supergravity. 
The dual gauge theory has 
%two phases in which the couplings of Yang-Mills theories differ from
%each othe% 
a different Yang-Mills coupling in each of the two halves of the
boundary spacetime divided by a codimension one defect. We discuss  
%the detailed shape of the geometry%
the geometry of our solution in detail, emphasizing the structure of
the boundary, and also study the string configurations corresponding
to Wilson loops. We also show that the background  is stable under
small scalar perturbations. 
%\end{titlepage}
\end{quote}
\vskip 1cm
\centerline{\today}
\newpage
\baselineskip 18pt

%%%%%%%%%%%%%%%%%%%%%%%%%%%%%%%%%%%%%%%%%%%%%%%%%
\def\nn{\nonumber}
%%%%%%%%%%%%%%%%%%%%%%%%%%%%%%%%%%%%%%%%%%%%%%%%%

\section{Introduction}
The AdS/CFT duality
\cite{Maldacena:1997re,Gubser:1998bc,Witten:1998qj}  beautifully 
exemplified the idea of
holography \cite{'tHooft:gx}\cite{Susskind:1994vu}  in clarity and
precision, more powerfully than any other 
examples of the gravity/field theory correspondence. Since
then there has already been a large volume of %, \lq\lq almost
%encyclopedic", 
applications of the AdS/CFT \cite{Aharony:1999ti}. 
Yet there may still be
%several 
novel interesting avenues to be unfolded and pursued in its
application.  

As an example, in this paper, we will consider a dilatonic deformation
of $AdS_5$ space\footnote{For other work on dilatonic deformations of
 AdS, see e.g. 
\cite{Kehagias:1999iy}\cite{Kehagias:1999tr}\cite{Gubser:1999pk}\cite{Girardello:1999hj}.}, i.e. an asymptotically $AdS_5$ space with a
spatially varying dilaton, which is a non-supersymmetric, yet
(classically) exact, solution of type IIB supergravity. In fact there
exists a class of dilatonic deformations of $AdS_5$ space of this
kind, where the dilaton varies not only spatially but also in
time. A time-dependent solution  might be interesting in its own
right. We will give a short description of these solutions in the
discussion.
However we found that the solution displayed a naked
singularity%at the center of AdS%
, and hence it is not clear whether
they are viable solutions in string theory.

Unlike the time-dependent solution, the deformation we will discuss in
this paper is regular everywhere, and breaks the isometry $SO(4,2)$
of $AdS_5$ space down to $SO(3,2)$. The spacetime is asymptotically
$AdS_5$, and the dilaton approaches to a constant at the
boundary. However the value of the constant dilaton differs in
one-half of the boundary from the other, leaving a codimension one
defect at their junction. Our dilatonic solution thus reveals two
different \lq \lq faces" at the boundary, which led us to a coinage 
the \lq\lq Janusian" solution.\footnote{It is dubbed after a Roman
  God, Janus, who has two faces.}   

%Having the $SO(3,2)$ isometry unbroken and the codimension one defect
%at the boundary, our deformation is 
%thought of as 
%somewhat similar to the
%the supergravity
%dual of a defect conformal field theory (dCFT). 
%The dCFT, in the
%context of the AdS/CFT, has already been studied 
%extensively 
%in
%\cite{Karch:2000gx,DeWolfe:2001pq,Bachas:2001vj,Aharony:2003qf}. 
%In this dCFT, the bulk source of the defect corresponds to branes
%in the AdS space.  In our solution, however,  there exist no bulk branes
%unlike the case of dCFT.
 
Being non-supersymmetric, 
our solution is potentially 
unstable. We will analyze and discuss the stability of  scalar
perturbations about the 
solution, a more complete analysis of the stability including other
modes is left for future work. From the dual field theory viewpoint,
we are turning on a 
marginal deformation by the operator ${\cal O}_4\sim\mbox{Tr}(F^2)$ dual to
the dilaton, but with different deformation parameters in different
halves of the flat boundary spacetime. The deformation is marginal, so
it preserves the dilation invariance, $SO(1,1)$. But it breaks a part of
the Lorentz symmetry down to $SO(2,1)$ due to the jump of the
deformation parameter at the codimension one defect. Hence 
the ($2+1$)-dimensional conformal invariance, $SO(3,2)$, is left unbroken.

Having the $SO(3,2)$ isometry unbroken and the codimension one defect
at the boundary,
%It may also be noted that 
our solution is quite reminiscent of that of
\cite{Karch:2000ct},
%\cite{Karch:2000gx,DeWolfe:2001pq,Bachas:2001vj,Aharony:2003qf}, 
where an $AdS_4$ brane inside an  $AdS_5$
space was considered and 
it was shown that gravity is locally localized on the $AdS_4$
brane.\footnote{See \cite{Porrati:2001gx}\cite{Porrati:2001db} for a
  holographic interpretation of the 
  Karch-Randall model in terms of 4-d gravity coupled to a conformal
  field theory. Note also that a UV brane in these models 
 need not be a true brane,  but specifies a UV-cutoff.} 
The dual gauge theory is a defect conformal field theory (dCFT), and in the
context of the AdS/CFT, it has already been studied 
%extensively 
in
\cite{Karch:2000gx,DeWolfe:2001pq,Bachas:2001vj,Aharony:2003qf}.
However our solution differs significantly from \cite{Karch:2000ct}  in that
we do not have a source $AdS_4$ brane at all. Further the detailed 
form of 
the gravity dual of dCFT has not been known thus far.
Our solution, though nonsupersymmetric, provides a simple example of
an explicit gravity dual of dCFT. We will make more 
comments on these points later. 

The organization of our paper is as follows. In section 2, we will
review a few different parameterizations of AdS space, and discuss
their boundary structure in some detail. In section 3, we will
present our dilatonic solution and discuss its properties. In section
4, we  will then discuss the geometry of the solution, in particular,
emphasizing  the structure of the boundary. We also make a comment on
a relation of our solution to that of \cite{Karch:2000ct}. In section
5, we will 
present the interpretation of our solution in terms of the dual $N=4$ SYM
theory, and also study the Wilson loop in our deformation. In section
6, we 
will discuss  the stability of our solution. We will end with a brief
discussion of  possible generalizations and a time dependent solution
which can be obtained by a slight modification of our ansatz.

\section{The coordinate systems} 
For later use, we will review a couple of different coordinate systems
for the $AdS_d$ space.  
The $AdS_d$ space is defined by a hyperboloid in $R^{2,d-1}$ 
\begin{equation}
X_0^2+X_d^2 - X_1^2 -\cdots -X_{d-1}^2 =1\ .
\label{adsone}
\end{equation}
\noindent
\underline{(I) The global coordinate}
\medskip

\noindent
The global coordinate covers the entire region of the $AdS_d$
space. It is obtained by the following parameterization,  
\begin{equation}
X_0= {\cos \tau\over \cos \theta},\quad  X_d = 
  {\sin\tau  \over \cos\theta}, \quad X_i = \tan\theta\,\, n_i ,\;\; 
  i=1,\cdots , d-1\ ,
\label{adstwo}
\end{equation}
where $n_i$ is the unit vector in $R^{d-1}$. The metric on the global  
$AdS_d$ is then given by 
\begin{equation}
ds^2_{AdS_d} = {1\over \cos^2\theta} \big( -d\tau^2 + 
  d\theta^2 +\sin^2\theta\, d\Omega_{d-2}^2\big)\ ,
\label{adsthree}
\end{equation}
with $\theta \in \left[0,{\pi\over 2}\right]$.

\medskip\noindent
\underline{(II) The Poincar\'e patch}
\medskip

\noindent
In the AdS/CFT, the Poincar\'e patch is the most convenient
parameterization of AdS space. Parameterizing $X_0={1 \over
  2}\left(z+\left(1+\vec{x}^2-t^2\right)/z\right)$, $X_d=t/z$,
$X_{i=1,\cdots,d-2}=x_i/z$, and $X_{d-1}={1 \over
  2}\left(z-\left(1-\vec{x}^2+t^2\right)/z\right)$, the metric takes
the form 
\begin{equation}
ds_{AdS_d}^2={1\over z^2}\left(-dt^2+d\vec{x}^2+dz^2\right)\ ,
\label{adsnine}
\end{equation}
where $\vec{x}=(x_1,\cdots,x_{d-2})$ and $z\in [0,\infty]$.

\medskip\noindent
\underline{(III) The $AdS_{d-1}$ slicing}
\medskip

\noindent
One can slice the $AdS_d$ space by  $AdS_{d-1}$ spaces, which corresponds to 
parameterizing $X_{d-1}=z$ and the rest, $X_{i=0,\cdots,d}$, by any
coordinate system of the $AdS_{d-1}$ space of radius
$\sqrt{1+z^2}$. Then the metric takes the following form 
\begin{equation}
ds_{AdS_d}^2= {dz^2\over 1+z^2} + (1+z^2) 
  ds^2_{AdS_{d-1}}\ .
\label{adsfour}
\end{equation}
Defining a new coordinate $\mu$ by $z=\tan\mu$, the metric is rewritten as
\begin{equation}
ds_{AdS_d}^2= f(\mu) \big(d\mu ^2 + ds^2_{AdS_{d-1} 
  }\big)\ ,
\label{adsfive}
\end{equation}
where $f(\mu)=1/\cos^2\mu$. Since the range of $z$ is  
$z\in[-\infty,+\infty]$, that of $\mu$ is $\mu \in \left[-{\pi\over 2}, 
{\pi\over 2}\right]$. This is  the most useful  coordinate system
for our application.

\medskip\noindent
\underline{The boundary geometry}
\medskip

\noindent
The conformal boundary of the global AdS metric (\ref{adsthree}) 
is located at $\theta =\pi/2$ and has the shape of $R\times S^3$ 
where $R$ is the time direction, while that of the Poincar\'e patch
(\ref{adsnine}) is at $z=0$ and of the shape $R\times R^3$. In the
$AdS_{d-1}$ slicing  
(\ref{adsfive}), the appearance of the boundary is less trivial. For 
later use, we will elaborate this point in some detail. 

First we take the global coordinate for the $AdS_{d-1}$ slice in
(\ref{adsfive}). Then the metric can be written as 
\begin{equation}
ds^2_{AdS_d} = {1\over \cos^2\mu\cos^2\lambda} 
 \big( -d\tau^2 + \cos^2\lambda d\mu^2 + 
  d\lambda^2 +\sin^2\lambda d\Omega_{d-3}^2\big)\ ,
\label{adssix}
\end{equation}
with $\lambda \in \left[0,{\pi\over 2}\right]$. The constant time
section of this metric is conformal to a half of $S^{d-1}$, since the
range of $\mu$ is   
from $-{\pi \over 2}$ to ${\pi \over 2}$. If $\mu$ had  ranged  
over $[-\pi,\pi]$, it would have been the full sphere. 
The boundary, $\mu=\pm\pi/2$, of this half sphere is exactly that of a
spatial section of  
the global $AdS_d$, thus being $S^{d-2}$. In fact the boundary consists of  
two parts, one of which is at $\mu=-\pi/2$ and the other at $\mu=\pi/2$. 
These two parts are joined through a surface, 
$\lambda=\pi/2$, which is a codimension one space in the boundary. 
The $\mu=\pi/2$ (or  $\mu=-\pi/2$ ) part is $H^{d-2}$, a half of
$S^{d-2}$, thus over all the boundary makes up the full $S^{d-2}$.  
Later we will encounter the case where the range of $\mu$ becomes 
larger. But the structure of the conformal boundary remains to be  
the same $S^{d-2}$. 

Next we take the Poincar\'e patch for the $AdS_{d-1}$ slice in
(\ref{adsfive}). Then the metric takes the form  
\begin{equation}
ds^2_{AdS_d} = {1\over y^2\cos^2\mu} 
 \big( -dt^2 + d\vec{x}_{d-3}^2 +dy^2 +y^2 d\mu^2\big)\ ,
\label{adsseven}
\end{equation}
with $y \in [0,\infty]$.  One can easily see that, 
by the change of coordinate $x=y\sin\mu$ and $z=y\cos\mu$, 
the above metric turns into the conventional form of  
the Poincare patch AdS,
\begin{equation}
ds^2_{AdS_d} = {1\over z^2} 
 \big( -dt^2 + d\vec{x}_{d-3}^2 +dx^2 + dz^2\big)\ .
\label{adseight}
\end{equation}
Again the boundary consists of two parts, one of which is at $\mu=\pi/2$ 
and the other at $\mu=-\pi/2$. These two parts are joined through a
codimension one surface, $y=0$, forming a $d-2$ dimensional flat
Euclidean space,  
 or $d-1$ dimensional Minkowski space when including the time. Even
with a larger range of $\mu$, as we will see below in our application,
the structure of the conformal boundary remains the same, on which the
dual $N=4$ SYM theory is defined.

\section{The ansatz and the Janusian solution} 
In the following we will consider a simple deformation of $AdS_5$
space, which is asymptotically $AdS_5$ and has a nontrivial dilaton
profile. We make the ansatz,  
\begin{eqnarray}
ds^2 &=& f(\mu) \left(d\mu^2 + ds^2_{AdS_{4}}\right)+ 
      ds_{S^5}^2\ , \nn\\
\phi&=&\phi(\mu)\ , \label{ansatza}\\
F_5&=& 2 f(\mu)^{5\over 2} d\mu \wedge \omega_{AdS_4}
+ 2 \omega_{S^5}\ ,\nn
\end{eqnarray}
where $\omega_{AdS_4}$ and $\omega_{S^5}$ are the unit volume forms 
on $AdS_4$ and $S^5$ respectively. Thus, in particular, the five
sphere $S^5$ is intact, so is the $SO(6)$ R-symmetry. But the
supersymmetry is completely broken. 
  
 The IIB supergravity equations of motion are given by
\begin{eqnarray}
&&R_{\alpha\beta} -{1\over 2}\partial_\alpha \phi 
\partial_\beta \phi 
  -{1\over 4} F^2_{\alpha\beta}=0\ ,\nn\\
&&\partial_\alpha(\sqrt{g} g^{\alpha\beta}\partial_\beta \phi)=0
\ ,\label{eqofm}\\ 
&&*F_5=F_5\ ,\nn
\end{eqnarray}
together with the Bianchi identity $dF_5=0$. 
The equation of motion for the dilaton (in the Einstein frame) is 
easily solved, giving 
\begin{equation}
\phi'(\mu) = {c_0\over f^{3\over 2}(\mu)}\ .
\label{dila}
\end{equation}
The Einstein equations give rise to
\begin{eqnarray}
2 f'f' -2 f f'' &=& -4 f^3 +{c_0^2\over 2} 
    {1\over f}\ ,\nn\\
12 f^2 + f'f' + 2 f f''&=& 16  f^3\,.
\label{einstein}
\end{eqnarray}
It is easy to see that these equations are equivalent to the first 
order differential equation 
\begin{equation}
f'f' = 4 f^3 -4f^2+{c_0^2\over 6} {1\over f}\ ,
\label{einsteinb}
\end{equation}
corresponding to the motion of a particle with zero  energy in a 
potential given by minus the right hand side (r.h.s.) of (\ref{einsteinb}).

\begin{figure}[ht!]
\centering \epsfysize=9cm
\includegraphics[scale=0.7]{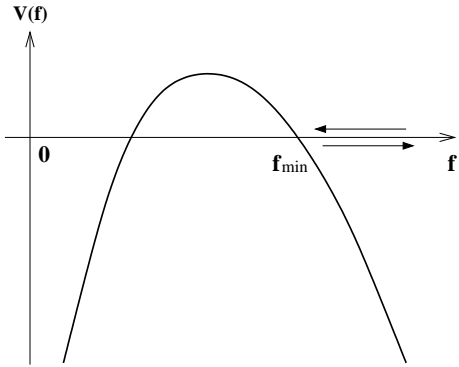}
\caption{\small The dynamics corresponds to the  
particle motion under a  potential with zero energy. We depict the
shape of the potential here. We are interested in the trajectory in which 
the particle starts from infinity, reflected at $f_{min}$ and goes back
to infinity.} 
\end{figure}

Equation (\ref{einsteinb}) can easily be integrated,
\begin{equation}
\mu %+\mu_0  
= \int^f_{f_{min}} {d\tilde{f} \over 2  \sqrt{\tilde{f}^3
    -\tilde{f}^2+{c_0^2\over 24}{1\over \tilde{f}}} }\ , 
\label{sola}
\end{equation}
where $f_{min}$ is the largest root of the denominator of the 
integrand. The function $f(\mu)$ is numerically solved in Figure 2.

\begin{figure}[ht!]
\centering \epsfysize=9cm
\includegraphics[scale=0.7]{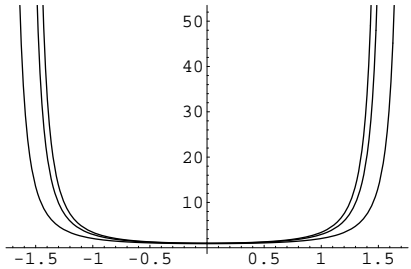}
\caption{\small $f(\mu)$ as a function of $\mu$ for values of $c_0= 
0.2, 0.8,1.3$}
\end{figure}

Note that $c_0=0$ 
corresponds to a constant dilaton and (\ref{sola}) gives the $AdS_4$ slicing 
of (\ref{adsfive}). The integration region of $f$ is bounded by the (simple)  
zero of the r.h.s. of (\ref{einsteinb}). Indeed the properties of the
solution are determined by the behavior of the rational function
\begin{equation}
p(x)= x^3-x^2+{c_0^2\over 24 x}\ ,
\label{pola}
\end{equation}
characterizing the potential, which
has the following properties: For $c_0\in [0, {9\over 4\sqrt{2}}]$, it 
has two real zeros lying between 0 and 1. 
For $c_{cr}= {9\over 4\sqrt{2}}$, two of the zeros coalesce and the 
function becomes
\begin{equation}
p(x)|_{c_0\to {9\over 4\sqrt{2}}} = {1\over x}\left(x-{3\over 
4}\right)^2\left(x^2+{1\over 2}x + {3\over 16}\right)\ .
\label{polb}
\end{equation}
For the critical $c_{cr}$ the integral (\ref{sola}) diverges
logarithmically.
For general $c_0\in [0, {9\over 4\sqrt{2}}]$ the rational function is of the 
form
\begin{equation}
p(x)= {1\over x} (x-x_1)(x-x_2) ((x-x_3)^2+ y_3^2)\ ,
\label{polc}
\end{equation}
where $x_{1,2,3}$ and $y_3$ are functions of $c_0$.

As we increase $c_0$, the integrand in (\ref{sola}) is monotonically
 decreasing, but the integration over $f$ grows so quickly that it
 offsets 
 and overwhelms the decreasing of the integrand.  
As shown in Figure 3 numerically,
the total range of $\mu$ that is 
covered by the solution --the total time it takes for the particle to 
move from infinity and back-- increases with increasing $c_0$.  
%As we take $\mu\in [-\mu_0,\mu_0]$, the total range will be $2\mu_0$. 
 
\begin{figure}[ht!]
\centering \epsfysize=9cm
\includegraphics[scale=0.7]{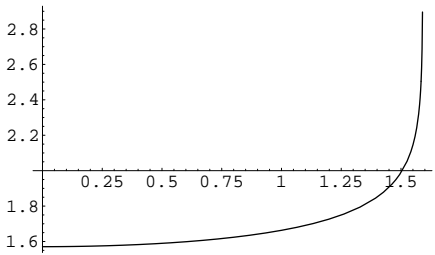}
\caption{\small The range of the coordinate $\mu$ as a functions of 
$c_0$, where $c_0$ varies from $0$ to $c_{crit}= {9\over 4\sqrt{2}}$, 
$\mu_0$ starts from 
$\mu_0={\pi\over 2}$ and diverges at $c_0=c_{crit}$ } 
\end{figure}

Note that, since $f\to \infty$ at the edge of the range of $\mu$, the 
dilaton approaches to a constant in this limit. Hence for the noncritical 
range of $c_0$ --as long as $c_0$ is smaller than the critical value
given below-- the dilaton varies over a finite range and the string
coupling can be made  
arbitrarily small;
\begin{eqnarray}
\phi(\mu_0)-\phi(-\mu_0) &=&  \int_{-\mu_0}^{\mu_0} 
{c_0d\mu \over 
    f^{3\over 2}(\mu) }\nn\\ 
&=& 2\int_{f_{min}}^\infty {c_0df \over 2 f^{3\over 2} \sqrt{f^3-f^2
+{c_0^2\over 24}{1\over f}}}\ ,
\label{dilb}
\end{eqnarray}
where $\pm\mu_0$ is the maximum/minimum value of $\mu$, i.e
$\mu\in [-\mu_0,\mu_0]$. The dilaton equation
(\ref{dila}) is numerically solved in Figure 4. Also the range of the
dilaton, as $c_0$ varies, is shown numerically in Figure 5.

\begin{figure}[ht!]
\centering \epsfysize=9cm
\includegraphics[scale=0.7]{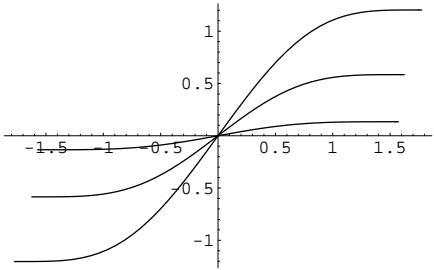}
\caption{\small 
The dilaton  $\phi$  as a functions of $\mu$, for values of $c_0= 
0.2,\; .8, \; 1.3$}
\end{figure}

\begin{figure}[ht!]
\centering \epsfysize=9cm
\includegraphics[scale=0.5]{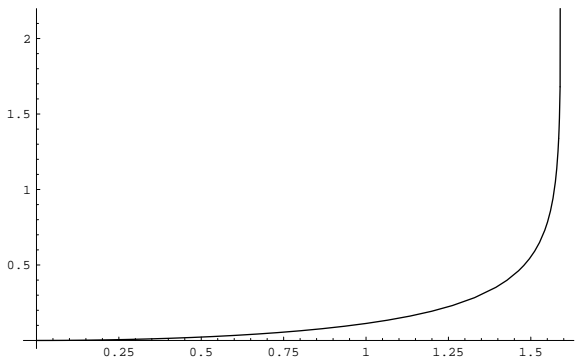}
\caption{\small The range of the dilaton  $\phi$ as a function of 
$c_0$, the range of the dilaton diverges at $c_0=c_{crit}$ } 
\end{figure}
 
\medskip 

Finally if $c_0$ exceeds $c_{cr}$, the zero energy motion of 
particle reaches the point $f=0$, where the geometry develops a 
naked singularity. We shall not consider this regime of upper 
critical solution below because we do not know how to deal with
the singularity. It is an interesting question why such a drastic 
transition
of geometry occurs depending on the strength $c_0$ of the 
dilaton perturbation at the boundary.

\section{The geometry of the Janusian solution} 
In this section we would like to describe the geometry 
of our solution with an emphasis on the structure of the boundary. 
Since there is no change in the $S^5$ part, we only look at the
$AdS_5$ directions. Adopting the global coordinate for the $AdS_4$
slice  
as in (\ref{adssix}), the metric becomes 
\begin{equation}
ds^2 = {f(\mu)\over\cos^2\lambda} 
 \big( -d\tau^2 + \cos^2\lambda d\mu^2 + 
  d\lambda^2 +\sin^2\lambda d\Omega_2^2\big)\ .
\label{geo}
\end{equation}
The spatial section of the conformal metric, i.e. the metric inside
the parenthesis is depicted in Figure 6.  
Only the surface of the globe parametrized by $\mu$ and $\lambda$ is
relevant here. Each point on the surface represents $S^2$ and the
boundary indicated by the bold lines is $S^3$.  
The boundary consists of two parts; one is at $\mu=-\mu_0$
and the other at $\mu=\mu_0$. These two halves of $S^3$ are joined
through the north and south poles. The dilaton varies from one constant, 
$\phi_0^-=\phi(-\mu_0)$, at one half of the boundary at $\mu=-\mu_0$,
to another, $\phi_0^+=\phi(\mu_0)$, at the other half of the boundary
at $\mu=\mu_0$, running through the bulk as $\mu$ changes. 
 
\begin{figure}[ht!]
\centering \epsfysize=9cm
\includegraphics[scale=0.7]{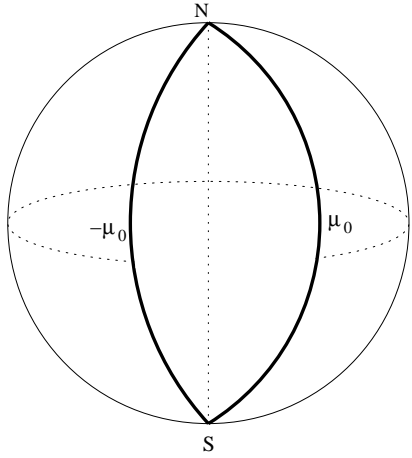}
\caption{\small The conformal mapping of the constant time slice 
 is depicted here. Only the surface 
of the globe described by the $\mu$ and $\lambda$ 
coordinates is relevant here. 
Each point on the surface corresponds to $S^2$. 
The boundary indicated by bold lines corresponds to $S^3$.}
\end{figure}

In the pure AdS case,
the geometry covers only a half of the surface of the globe; the
spatial section of the metric (\ref{adssix}) is conformal to  
$R\times H^3$.
As shown in the last section, the range of $\mu$ becomes larger as
$c_0$ grows. There is a value of $c_0$ when the range of $\mu$ becomes
$\mu \in [-n\pi,n\pi]$ with $n$ being integer, thus the geometry
(multiply) covering up the entirety of the globe. But we do still have
the boundary, as we do not identify two halves of the boundary 
--if two halves of the boundary were identified, 
the spatial section of the geometry would have become (a multi-cover
of) $S^4$ without boundary. Hence the boundary always consists of two
halves of $S^3$, independent of the value of $c_0$ or equivalently of
$\mu_0$. As $c_0$ approaches its critical value $9/(4\sqrt{2})$, the
range of $\mu$ extends to the entire real line. 

When we adopt the Poincar\'e patch for the $AdS_4$ slice as in
 (\ref{adsseven}), 
 the metric is written as
\begin{equation}
ds^2 = {f(\mu)\over y^2} 
 \big(-dt^2 + d\vec{x}^2 +dy^2 +y^2 d\mu^2
\big)\ .
\label{geoa}
\end{equation}
The conformal mapping of the spatial section is depicted in Figure 7.
Each point on the plane represents ${ R}^2$.
As in the pure AdS case, $y=\infty$ corresponds to the horizon.
Again the boundary is at $\mu=\pm \mu_0$. Each of these is a half of
${R}^3$, being joined together through the wedge $W$ that is 
${ R}^2$. The range of $\mu$ is the same as before; it becomes larger
as $c_0$ grows.

\begin{figure}[ht!]
\centering \epsfysize=9cm
\includegraphics[scale=0.7]{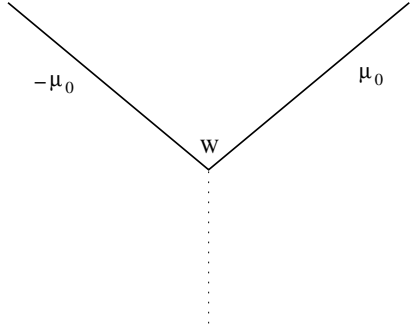}
\caption{\small The conformal diagram of the spatial section
of our solutions is depicted here for the Poincare type coordinate. 
Each point on the plane corresponds to ${ R}^2$.
The coordinate $\mu$ is the polar angle and $y$ is the radial coordinate
with $W$ as the origin.   
The total boundary consists of two parts, each of which has the geometry of
a half of ${R}^3$.  These two parts are joined through $W$, which may be 
viewed as a kind of  domain wall on the boundary. } 
\end{figure}

Again the dilaton varies from one constant, $\phi_0^-=\phi(-\mu_0)$,
at one half of the boundary $\mu=-\mu_0$, to another,
$\phi_0^+=\phi(\mu_0)$, at the other half $\mu=\mu_0$,  running
through the bulk as $\mu$ increases.  
Hence from the viewpoint of the boundary, the dilaton is constant
in each half, taking the value of $\phi_0^+$ and $\phi_0^-$
respectively. 
The discontinuity of the dilaton occurs through the joint $W$. 
One might worry about possible singularities around the wedge
$W$. However, this is just an artifact of the conformal diagram. That
is,  
\ from the viewpoint of the original geometry, nothing singular happens around 
the wedge. The same conclusion follows from the stability analysis as we will 
see below.

\subsection{A comparison to Karch-Randall} 
In \cite{Karch:2000ct} Karch and Randall considered the issue of local
localization of gravity on a brane in AdS. In particular they studied
$AdS_4$ branes  
in $AdS_5$. The  geometric picture of their solutions --two `halves' 
of 
$AdS_5$ glued together along a $AdS_4$ brane is quite similar to 
ours. Note however the differences: We do not have a singular brane 
source in the bulk. The discontinuity only appears on the boundary. 

The coordinate change going from the $AdS_4$ slicing 
\begin{equation}
ds^2_{AdS_5} = {1\over \cos^2\mu}\Big( d\mu^2 + {1\over 
\cos^2\theta}\big (-dt^2+d\theta^2+\sin^2\theta
d\Omega_{2}^2\big)\Big) 
\label{adsla}
\end{equation}
\medskip 
to the standard global $AdS_5$ coordinates 
\begin{equation}
ds^2_{AdS_5}= -\cosh^2\tilde{r} dt^2 + d\tilde r^2+ 
\sinh^2\tilde{r} ( d\rho^2 +\sin^2\rho d\Omega_{2}^2)\ ,
\label{adsslb}
\end{equation}
is given by
\begin{equation}
\tan%\tilde
{\rho}= {\tan\theta\over \sin\mu}\ , \quad 
\quad \cosh\tilde{r} ={1\over \cos\mu\cos\theta}\ ,
\label{coordch}
\end{equation}
where the range of the coordinates is as follows; For (\ref{adsla}) one 
finds $\mu\in[-\pi/2, \pi/2]$ and $ \theta\in[0,\pi/2]$, for (\ref{adsslb}) 
one has $\tilde r\in[ 0,\infty]$ and $\rho\in[0,\pi]$.   
This coordinate system was  used to describe the $AdS_4$ brane
in \cite{Karch:2000ct}.

%\medskip 
 
In \cite{DeWolfe:1999cp} five dimensional gravity coupled to a scalar
$\phi$ with  
potential $V(\phi)$ and a localized co-dimension one brane source, was
studied.   
In section 3.2 the case of an AdS worldvolume of the brane source was 
studied corresponding to a $AdS_4$ slicing of $AdS_5$. The dilaton 
would be a special case where the potential is simply the cosmological 
constant $V(\phi)=\Lambda$, without any dependence on the dilaton.
Note that in our solution there is no singular brane source where
the value of fields jump, the solution is more like a (very thick)
domain wall.

\section{The holographic dual}
The solution given by (\ref{ansatza}), (\ref{dila}), and (\ref{sola})
is very simple, since only the dilaton and the metric
develop a nontrivial profile. In this section we will discuss the
holographic interpretation of this solution in terms of the theory
living on the boundary. The solution breaks the $SO(4,2)$ $AdS_5$
symmetry to $SO(3,2)$. Note that this situation is very similar to the
$AdS_{4}$ brane in $AdS_{5}$ which displays local localization of
gravity \cite{Karch:2000ct} and its relation to defect CFT
\cite{Karch:2000gx}\cite{DeWolfe:2001pq}. There the 
reduced AdS 
symmetry is interpreted as the part of the conformal symmetry group
which leaves the defect invariant.

Near the boundary $\mu\to \pm \mu_0$, one can easily  show that $f(
\mu)$ and the dilaton $\phi(\mu)$ behave as follows;
\begin{equation}
f= {1\over  (\mu\mp\mu_0)^2}+o(1)\ ,\quad \quad \phi= \phi_0^\pm \mp
{c_0\over 4}(\mu\mp\mu_0)^4 + o((\mu\mp \mu_0)^6)\label{expanb}\ . 
\end{equation}
In the standard AdS/CFT  dictionary the asymptotic value of the dilaton is
identified with the coupling constant of the four dimensional $N=4$ SYM
theory living on the boundary.   The subleading term is identified
with the expectation value for the operator $O_4=\mbox{Tr}(F^2)$. Hence the
interpretation of our solution is that the boundary consists of two
half spaces, given by $\mu= \pm \mu_0$, where the coupling constant is
given by 
\begin{equation}
\mu=+\mu_0: \quad {g_{YM}^2\over 4 \pi}= e^{ \phi_0^+}, \quad \quad
\mu=-\mu_0: \quad {g_{YM}^2\over 4 \pi}= e^{ \phi_0^-}\ . 
\end{equation} 
Note however that in our solution we do not need either a potential
for the scalar field or a singular brane source in the bulk as in
\cite{DeWolfe:1999cp}.
 
\subsection{Wilson Loops}
In \cite{Maldacena:1998im} it was proposed that the calculation of 
the expectation value $\langle W(C)\rangle$ of Wilson loop operators  in the
large $N$ SYM theory,  
\begin{equation}
W(C)= {1\over N} {\rm tr} P e^{i \oint_C A}\ , 
\end{equation}
have a dual holographic interpretation as the
minimal 
area of a 
fundamental string worldsheet in the AdS bulk, which ends on the boundary.
The action for the fundamental string is given by
\begin{equation}
S= -{1\over 2 \pi \alpha'}\int d\tau d\sigma 
\sqrt{-det(g_{\mu\nu}\partial_a X^\mu \partial_b X^\nu)}\label{acwila}\ .
\end{equation}
Where   $g_{\mu\nu}$ is the string frame metric. We assume that the
     Wilson loop is static and choose the temporal gauge
$\tau=X^0$, furthermore we assume that the Wilson loop will not depend
     on the coordinates $x_i$. Plugging in the form 
of the metric (\ref{geoa}) one gets for the Lagrangian.
\begin{equation}
L= -\int d \sigma e^{\phi\over 2} {f(\mu)\over y^2}\sqrt{ \partial_\sigma y
\partial_\sigma y + y^2 \partial_\sigma \mu \partial_\sigma
\mu}\label{acwilb}\ . 
\end{equation}
One can use the remaining reparameterization invariance to choose the
static gauge $\mu=\sigma$. The equation of motion following from
(\ref{acwilb}) will be
\begin{equation}
y''+ y +y'\left( {f'\over f}+{1\over 2}\phi'\right) 
\left( 1+ {(y')^2\over
y^2}\right)=0\label{eqwil}\ .
\end{equation}
Hence the Wilson loop is determined by a set of coupled ordinary
differential equations given by (\ref{eqwil}) together with
(\ref{dila}) and (\ref{einsteinb}). 
The value of the expectation value of the Wilson loop operator is
given by $e^{-S}$ for the solution. Using the
expansion near the boundary (\ref{expanb}), it is easy to see that
the action  diverges as
\begin{equation}
S\sim e^{\phi_0\over 2}|y_0| {1\over ( \mu\mp\mu_0)^2}+ o(1)\ .
\end{equation}
This divergence is exactly that of the Wilson loop in AdS
\cite{Maldacena:1998im} and can be regulated by subtracting the
(infinite) action of a straight Wilson loop going from the  boundary
to the horizon.

\begin{figure}[ht!]
\centering \epsfysize=9cm
\includegraphics[scale=0.7]{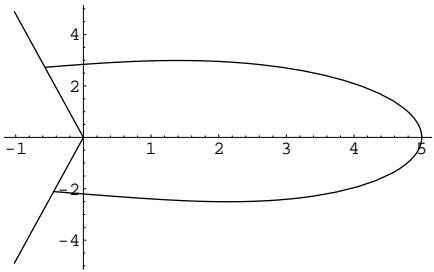}
\caption{\small Wilson line with $c_0=1.3$, and $y_{max}=5$.} 
\end{figure}

Unfortunately the equation (\ref{eqwil}) is too complicated to solve
analytically, however numerical solutions are easily obtained. The
main feature is that the Wilson loop will be tilted by the nontrivial
dilaton profile and metric warping, as can be seen from Figure 8.

\section{The stability against small perturbations} 
Our solution breaks all the supersymmetries of the IIB supergravity.
This fact  may be checked explicitly
by considering the supersymmetry variation of the dilatino and gravitino. 
Since there is no supersymmetry left unbroken, 
the background could potentially be unstable.
In order to check the
stability of our solution against small fluctuations, we shall 
 consider the behavior
of the scalar field in the background of our solution. 
The analysis may be generalized
to the metric or R-R field perturbations, but, for simplicity, 
we shall focus on the
case of the scalar field.
The stability of a solution requires that  
 there is no configuration 
with the same or smaller energy into which it can decay.
Here we are particularly interested in the 
perturbative stability of 
the background and will ask about the following two points: 

\noindent 1) The energy is positive definite for any modes of finite energy  
satisfying zero momentum flux condition at the boundary.
 The latter condition i.e. $\int dS_a\, T^a\!\,_0|_{\rm boundary}=0$ 
is required for the conservation of the  energy.

\noindent 2) One  requires the completeness of the modes   
to allow arbitrary initial configurations of small fluctuation.
 
In pure AdS space, it is well-known that not too tachyonic scalars  are 
perturbatively stable   unlike the case of the flat space. In $AdS_5$
 the scalar field is stable if the Breitenlohner-Freedman bound is
 satisfied, $m^2 \ge -4$. 
This is the question about the stability of the scalar
field itself and the issue here is different from the stability of 
the background. However as the methods are closely related, we would like 
to consider the generic massive 
scalars as well.  The $m^2=0$ case  will then correspond to the
stability of the  
background.  

We consider small fluctuations of the scalar field in the background 
of our solution. Specifically, we work in the 
Poincare type coordinate, 
\begin{equation}
ds^2 = {f(\mu)\over y^2} 
 \big( -dt^2 + d\vec{x}^2 +dy^2 +y^2 d\mu^2\big)\ ,
 \label{poincare}
\end{equation}
and  consider  the equation of motion
\begin{equation}
\left({1\over \sqrt{g}}\partial_\alpha \sqrt{g}g^{\alpha\beta}
\partial_\beta -m^2\right)\varphi=0\ ,
\label{massscalar}
\end{equation}
for the massive scalar field. Since any inhomogeneous fluctuations 
in $\vec{x}$
costs more energy, we consider only $\vec{x}$-independent modes. For
$\varphi= {\rm Re} 
(e^{-i\omega t} H (\mu, y))$, one finds that 
the spatial part $H$ satisfies
\begin{equation}
y^2 \partial_y y^{-2} \partial_y H + {1\over y^2}\left( f^{-{3\over2}}(\mu) 
\partial_\mu f^{{3\over 2}}(\mu)\,\partial_\mu -m^2f\right)H= -\omega^2 H\ .
\label{scalar}
\end{equation}
We solve this by the separation of  variables,
\begin{equation}
H = F(\mu) G(y)\ ,
\label{separation}
\end{equation}
with % and find a Bessel equation for 
F($\mu$) % which depends on the eigenvalue of the 
%$\mu$ equation,
satisfying
\begin{equation}
\left(-f^{-{3\over2}}(\mu)
\partial_\mu \, f^{{3\over 2}}(\mu)\,\partial_\mu + m^2f \right)
F(\mu)=k F(\mu)
\ .
\label{gegen}
\end{equation}
The function $G(y)$ satisfies 
\begin{equation}
G'' - {2\over y } G'  +(\omega^2 -k/y^2) G=0\ ,
\label{bessel}
\end{equation}
which may be solved by
\begin{equation}
G= y^{3/2} Z_\nu (\omega y)\ ,
\label{soly}
\end{equation}
where $\nu^2=9/4 + k$ and $Z_\nu (x)$ is the Bessel function.
For real $\omega$,  $Z_\nu$ may be either $J_\nu$ or $N_\mu$.
When $\omega$ is pure imaginary, we consider  only $Z_\nu=K_\nu (|\omega| y)$
because the other independent solution $I_\nu(|\omega|y)$ blows up
exponentially  at  
large $y$.

The energy of the system is defined by
\begin{equation}
E= {1 \over 2}\int dx^a \sqrt{g}\left( |g^{00}|\dot{\varphi}^2
+ g^{ab}\partial_a\varphi \partial_b\varphi + m^2 \varphi^2\right)\ .
\label{energy}
\end{equation}
Depending on the couplings of the scalar fields to the gravity, there may 
be a contribution of  
the improved term \cite{freedman} to the energy, but here we set this  
to zero. The inclusion of this contribution will not change the 
conclusion of our analysis.

For the mode given above, one may formally manipulate the energy 
functional
 using integration by parts with respect to $\mu$ 
and Eq.(\ref{gegen}). In the integration by parts, there may be a boundary 
contribution, but let us  assume that 
the boundary contribution vanishes. The energy functional then 
becomes
\begin{equation}
E= {1 \over 2}\int dx^a \sqrt{g} g^{yy} 
\left( \dot{\varphi}^2
+(\partial_y\varphi)^2  + {k\over y^2} \varphi^2 \right)
\ .
\label{energya}
\end{equation}
When $k+ 9/4 > 0$, the second and third term
give divergent contributions at small $y$ and the infinities do not 
cancel if  $\varphi \sim y^{3/2}N_\nu$ or $y^{3/2}K_\nu$. 
Hence only $\varphi \sim y^{3/2} J_\nu(\omega y)$ is allowed by the finiteness
when $k+ 9/4 > 0$. To show the positivity,  we rewrite the energy integral
 by $S$ defined by  $\varphi=y^{3/2 +\nu} S$. With  a little manipulation,
the expression for the energy becomes
\begin{equation}
E= {1 \over 2}\int dx^a \sqrt{g} g^{yy} y^{3+2\nu} 
\left( \dot{S}^2
+(\partial_y S)^2 \right)
\ ,
\label{energyb}
\end{equation}
for  $\varphi \sim y^{3/2} J_\nu(\omega y)$. Therefore we conclude
that it is sufficient that $k>-9/4$ for the energy of this mode to be
positive definite. 
 
Now let us consider the $\mu$-direction where we are interested 
in the integral
\begin{equation}
I= \int^{\mu_0}_{-\mu_0} d\mu \, f^{{3\over 2}} 
\left((\partial_\mu F)^2  + {m^2 f} F^2 \right)
\ ,
\label{angint}
\end{equation}
that is the last term of the energy integral (\ref{energya}).  
Setting $W=F f^{3/4}$, Eq.(\ref{gegen}) becomes
\begin{equation}
[-\partial^2_\mu  +V(f)]W(\mu)=k W(\mu)\ ,
\label{gegentwo}
\end{equation}
with
\begin{equation}
V= {3\over 4 f} \left(f'' -{1\over 4 f} f' f'\right) + m^2f
={3\over 4} \left(5f -3 - {c_0^2\over 8 f^3}\right) + m^2 f
\ ,
\label{port}
\end{equation}
where, for the second equality, we used (\ref{einsteinb}).
We now analyze the equation (\ref{gegen}) around $\mu \mp \mu_0\sim 0$,
 and there we have $$F= (\mu\mp\mu_0)^\gamma (1+{ h.\,o.\, t.})\,.$$ 
Since
$f\sim (\mu\mp\mu_0)^{-2} (1+{ h.\, o.\, t.})$, one finds
$\gamma (\gamma -4)=  m^2$ and its solutions are
$$ \gamma_\pm= {2}\pm \sqrt{4+m^2}\,.$$
\ From (\ref{angint}), we see that 
only $\gamma_+$ makes the integral finite when $m^2 +4 \ge 0$.

Let us first consider  the $c_0=0$ case
 corresponding to the pure AdS. When $m^2 +4 \ge 0$,
with $K=  (f^{1\over 2})^{2+\sqrt{4+m^2}}F$,  the integral $I$ 
can be rewritten as
\begin{equation}
I= \int^{\pi\over 2}_{-{\pi\over 2}} d\mu \, 
f^{-\tilde{\gamma}}  \left[(K')^2 + (\tilde{\gamma}^2
-9/4) K^2\right] > {-{9\over 4}}
\int^{\pi\over 2}_{-{\pi\over 2}} d\mu \, 
f^{-\tilde{\gamma}}  K^2\ ,
\label{anginta}
\end{equation}
where $\tilde\gamma=1/2 + \sqrt{m^2+4}$.
This can be shown by the integration by parts, in which
the boundary contributions vanish. On the other hand, 
using the equation (\ref{gegen}) 
and integration by parts, one may also be able to show that 
\begin{equation}
I= k\int^{\pi\over 2}_{-{\pi\over 2}} d\mu \, 
f^{-\tilde{\gamma}}  K^2\ .
\label{angintb}
\end{equation}
Thus we conclude that, for $c_0=0$, the condition $k > -9/4$ is not
only sufficient but also necessary for the energy of this mode to be
finite and positive definite. This way we have shown that the
condition $m^2+4 \ge 0$ is sufficient for the energy to be finite and
positive definite. 
On the other hand, if $m^2 +4 <0$, there may be a mode that has finite energy, 
but there is no mode satisfying zero momentum flux condition at the boundary
$\mu=\pm \mu_0$. Then there is no way to satisfy the completeness 
condition of the modes. In fact, the stability is not about the modes but 
of an arbitrary initial configuration of small fluctuations. 
The fact that there is no way to respect the zero flux condition at
the boundary in the subsequent time evolution clearly illustrates the
instability.

As $c_0$ grows, the potential $V(f)$ becomes smaller and 
the interval $2\mu_0$ gets larger. The latter makes the kinetic energy
contribution generically smaller. Hence one expects that
the mass bound $m^2$ will grow as $c_0$ gets larger. Unfortunately,
the evaluation of the precise bound as a function of $c_0$ is not trivial.
Hence we will consider the case $c_0\rightarrow 9/(4\sqrt{2})$.
In this case, we note that the integral  $I$ may be written as
\begin{equation}
I=\int^{\mu_0}_{-\mu_0} d\mu \left((W')^2 +V W^2 \right)
=k\int^{\mu_0}_{-\mu_0} d\mu\, W^2\ ,
\label{integral}
\end{equation}
for the finite energy modes.
Since $\mu_0\rightarrow \infty$ the kinetic contribution in 
(\ref{integral}) can be made arbitrarily small, while making full use
of the region where the potential becomes minimal. Namely if one allows 
an infinitesimal deviation from the minimum, the corresponding range
in $\mu$ becomes very large. Hence the evaluation of mass bound 
follows from the condition that the minimum of the potential
should be equal to or larger than $-9/4$. 
When $m^2 < -15/4$, the potential is unbounded from below, so we restrict
our attention to the case when $m^2 \ge  -15/4$.
Then the minimum of the potential is 
$$V_{min}= {3(3+m^2)\over 4} -9/4$$
where we have used the fact that the minimum $f_{min}$ of $f$ is given by $3/4$
for $c_0= 9/(4\sqrt{2})$.
This leads to the stability condition
$m^2 \ge -3$ for $c_0= 9/(4\sqrt{2})$.

\ From the above analysis, the stability of the background (i.e. 
the case of $m^2=0$) 
is quite clear. As a by-product, it is convincing that nothing singular
happens around $y=0$ contrary to the first impression of the geometrical shape.

In the discussion of stability above, we omitted many details. 
We do not consider the metric or R-R field  perturbations.
For the scalar field, we do not include the case where
there is the improved term for the energy. As we said earlier, 
this inclusion will not change our result of stability analysis. 
We  do not present detailed discussion of the zero flux condition, 
the role of completeness of mode and so on. 
We leave these issues to the further studies. 

\section{Discussion}
 In this note we have presented a very simple deformation of
 the $AdS_5\times S^5$ background for type IIB string theory. The main
 property of the solution is that the AdS space gets `elongated' 
and
 the dilaton develops a nontrivial profile. In terms of the
 representation of $AdS_d$ as a hyperboloid in $R^{2,d-1}$ given by
 (\ref{adsone}), the ansatz for the solution simply takes the dilaton
 to be a function of one of the $X_i, I=1,\cdots,d-1$. This
 construction can obviously be generalized to AdS spaces of different
 dimensionality as long as  massless scalar fields are present.

 This immediately
 suggests a generalization of our solution which makes the dilaton a
 function of one of the timelike coordinates $X_0$.
The ansatz for the metric, dilaton and five-form is given by
\begin{eqnarray}
ds^2&=&{f(X_0)\over |1-X_0^2|}ds_{AdS_5}^2+ds_{S^5}^2\ ,\\
\phi&=&\phi(X_0)\ ,\\
F_5&=&4\left({f(X_0)^{5/2}\over |1-X_0^2|^{5/2}}\omega_{AdS_5}
+\omega_{S^5}\right)\ .
\end{eqnarray}
The equation of motion determines $f(X_0)$
\begin{eqnarray}
\phi'(X_0)&=&c_0|1-X_0^2|^{1/2}f^{-3/2}\ ,\\
(1-X_0^2)^2f'^2&=&4f^2-4{1-X_0^2\over |1-X_0^2|}f^3
+{c_0^2\over 6}f^{-1}\ .
\end{eqnarray}
For $c_0\neq 0$ there are two branches of the solution with $X_0^2>1$
and $X_0^2<1$. 
Note that in contrast to to the spacelike case discussed in section 3, 
$f$ reaches zero in finite time.  
Since the scalar curvature is given by
\begin{equation}
R=-20-{c_0^2\over 2f^4}\ .
\end{equation}
This implies that this solution has  naked  singularities.
For the branch  $X_0^2>1$ this is a timelike singularity whereas for
$X_0^2<1$ it is a spacelike singularity. Note also that in the second
case $f$ does not reach infinity and hence there is no boundary.

 It is an open and very interesting question whether
 this singularity can be resolved or given a proper interpretation (as
 an unstable decaying brane and a cosmology respectively) in
 the full string theory.

Finally note that,
in finding our gravity solution, 
we make use of the unbroken $SO(3,2)$ invariance
to simplify the equations of motion. However, the ansatz we used is
not the most general one consistent with the $SO(3,2)$ symmetry.
Any generalization of our ansatz with the symmetry would be quite 
interesting. In particular, this may be helpful in 
finding the supergravity 
solutions of the $AdS_4\times S^2$ branes in $AdS_5\times S^5$
with inclusion of full back reaction.

\section*{Acknowledgments}
 We are grateful to Carlos Herdeiro, Andreas Karch, 
Per Kraus, and Joan Simon for useful discussions and conversations.
DB and SH would like to thank the warm hospitality of ITP Stanford University
where part of this work was done.
The work of MG was supported in part by NSF grant 9870115.
The work of DB is supported in part by KOSEF 1998     
Interdisciplinary Research Grant 98-07-02-07-01-5. 
The work of SH was supported in part by Israel Science Foundation
under grant No. 101/01-1.

\end{document}